\documentclass[conference]{IEEEtran}
\IEEEoverridecommandlockouts
\usepackage[colorlinks,linkcolor=blue,anchorcolor=blue,citecolor=blue]{hyperref}
\usepackage{cite}
\usepackage{url}
\usepackage[caption=false]{subfig}
\usepackage{float}
\usepackage{makecell}
\usepackage{amsmath,amssymb,amsfonts}
\usepackage{hyperref}
\allowdisplaybreaks[4]
\usepackage{graphicx}
\usepackage{enumerate}
\usepackage{array}
\usepackage{multirow}
\usepackage{color}
\usepackage{colortbl}
\usepackage[ruled,vlined,linesnumbered]{algorithm2e}
\usepackage{tikz}

\def\BibTeX{{\rm B\kern-.05em{\sc i\kern-.025em b}\kern-.08em
    T\kern-.1667em\lower.7ex\hbox{E}\kern-.125emX}}
\begin{document}

\title{Phase-Coordinated Model Predictive Control for AI Training Power Smoothing\\

\thanks{This work was supported by the National Key Research and Development Program under Grant 2024YFB4206500.}
}

\author{\IEEEauthorblockN{Jie Zhu, Ye Tian, and Yinliang Xu*}
\IEEEauthorblockA{\textit{Shenzhen International Graduate School} \\
\textit{Tsinghua University}\\
Shenzhen, China \\
zhuj24@mails.tsinghua.edu.cn, tiany25@mails.tsinghua.edu.cn, xu.yinliang@sz.tsinghua.edu.cn}

}

\maketitle

\begin{abstract}
Periodic power fluctuations from artificial intelligence (AI) training workloads can excite oscillations in power systems. This paper proposes a model predictive control (MPC) framework that exploits short start-time delays to mitigate power fluctuations at the computing source. By coordinating the relative phases of independent training jobs, the proposed method reduces the superposition of oscillatory power components without interrupting running jobs or requiring additional power-smoothing hardware. To obtain a tractable optimization formulation, cosine and sine projections are employed to quantify the dominant oscillatory power components, and their magnitudes are used to construct a convex quadratic power-smoothing objective. The receding-horizon framework updates launch decisions as new workload information becomes available while maintaining scheduling feasibility under limited computation time. Numerical results demonstrate reductions of 74.39\% in mean-square power fluctuations within the 0.2--2-Hz band and 46.6\% in power standard deviation, with an average launch delay of 2.188~s. All launch deadlines are satisfied, and the maximum end-to-end computation time is 0.390~s, below the 0.50-s MPC update period.
\end{abstract}


\section{Introduction}
The expansion of AI infrastructure is increasing both the magnitude and concentration of data center (DC) electricity demand. The International Energy Agency estimates that DCs consumed approximately 415~TWh in 2024 and projects about 945~TWh in 2030 under its base-case scenario \cite{iea2025energy}. Accommodating this growth requires coordination between computing operations and the power system. Existing studies exploit workload flexibility for demand response \cite{sun2025energy} and carbon-aware scheduling \cite{radovanovic2023carbon}. These approaches establish the value of adjusting computing demand in response to economic and environmental signals.

AI training also creates a distinct dynamic challenge. Repeated transitions between computation-intensive and communication-intensive phases can produce substantial, sustained power fluctuations \cite{song2026oscillation}. When their frequency components overlap with weakly damped grid modes, these variations can act as forcing inputs and amplify local or inter-area oscillations \cite{azizi2025grid}. Managing average energy consumption alone does not address this mechanism. Recent studies have developed dynamic data-center models for power-system stability analysis \cite{gyang2026dynamic} and investigated power-system oscillations induced by AI workload fluctuations \cite{11487948}. These findings highlight the need for operational strategies that explicitly account for the temporal characteristics of computing demand.

Energy storage can mitigate workload-induced power fluctuations by compensating rapid variations in DC demand. Battery energy storage systems (BESSs) and supercapacitors (SCs) provide complementary power and energy capabilities for smoothing fluctuations across multiple timescales. Hybrid storage has been investigated for cost-efficient power shaving and fluctuation mitigation \cite{11581359,ko2026mitigation}. More recently, Chen proposed a grid-mode-aware model predictive control (MPC) framework that coordinates storage dispatch to suppress oscillatory power components near vulnerable grid modes \cite{chen2026gridmode}. However, storage-based smoothing may require additional investment, while repeated charging and discharging can accelerate battery degradation. Furthermore, storage resources are also needed for backup power and grid services, creating competing demands for their limited capacity.

Workload scheduling offers an alternative means of regulating DC demand by adjusting when and where computing tasks are executed. Recent studies have exploited spatiotemporal flexibility to coordinate DCs across regional electricity markets \cite{jin2025spatiotemporal} and integrate day-ahead planning with real-time job placement for carbon-aware operation \cite{hall2025carbon}. These approaches primarily focus on energy management over relatively long scheduling horizons rather than short-term power fluctuations. Moreover, extended workload shifting may increase job completion times, introducing trade-offs between economic benefits and service quality \cite{sun2025power}.

In contrast, short start-time adjustments offer an opportunity to mitigate periodic power fluctuations with minimal impact on computing performance. For AI training jobs lasting several hours, a launch delay of only a few seconds represents a small fraction of the total execution time but can substantially change the relative phases of their periodic power components. Coordinating these delays can reduce the constructive superposition of workload fluctuations and smooth aggregate demand without interrupting running jobs or requiring additional energy storage equipment.

Motivated by this observation, this paper proposes a model predictive control (MPC)-based start-time scheduling framework that coordinates the launch times of waiting AI training jobs. The framework models aggregate DC power as the combination of fixed power profiles from running and committed jobs and time-shifted profiles of waiting jobs. Dominant oscillatory components are characterized through cosine and sine projections, whose squared magnitudes form a tractable convex quadratic power-smoothing objective. The resulting optimization determines suitable launch delays to mitigate aggregate power fluctuations, while the receding-horizon framework updates scheduling decisions as new workload information becomes available.

\section{Workload Modeling and Scheduling}
\label{sec:model and Scheduling}
This section models how job start times change the temporal alignment and aggregate power of AI workloads.

\subsection{ MPC-based Start-Time Scheduling Framework }
\label{sec: MPC_framework}
The scheduler coordinates independent, delay-tolerant jobs. Each job follows its original power profile continuously after launch; only its start time is adjustable. Three time scales coordinate power prediction and online scheduling. First, $\Delta_p$ denotes the power-sampling interval used to discretize predicted workload profiles and extract their dominant frequency components. A smaller $\Delta_p$ enables finer characterization of rapid power fluctuations. Second, $\Delta_d$ specifies the spacing between candidate launch times, determining the resolution of start-time decisions and phase adjustments. Third, $T_c$ denotes the MPC update period, with updates at $t_k=kT_c$. At each update, the controller refreshes predictions and reoptimizes waiting-job launches, committing only near-term decisions. The ratios $\Delta_d/\Delta_p$ and $T_c/\Delta_d$ are chosen as positive integers to ensure temporal alignment among power sampling, job launches, and MPC updates.

At each MPC update $t_k$, scheduled launches and job completions are first processed. The jobs are then classified into three disjoint sets: $\mathcal R_k$ contains running jobs, $\mathcal C_k$ contains jobs with previously committed launches in $(t_k,t_{k+1})$, and $\mathcal J_k$ contains arrived jobs whose launch times remain adjustable. The launch times of jobs in $\mathcal R_k$ and $\mathcal C_k$ are fixed, while only those in $\mathcal J_k$ are optimized at the current update.

Job $j$ arrives at $a_j$, executes for $L_j$, and may wait at most $\bar d_j$. Its deadline $\ell_j=a_j+\bar d_j$ remains fixed across updates. The launch time $s_j$ and waiting time $d_j$ satisfy
\begin{equation}
t_{k+1}\leq s_j\leq\ell_j,\quad
d_j=s_j-a_j,\quad \forall j\in\mathcal J_k.
\label{eq:launch}
\end{equation}

The lower bound reserves $[t_k,t_{k+1})$ for optimization and dispatch. Launches in $[t_{k+1},t_{k+2})$ are then committed, while later decisions remain adjustable. To accommodate arrivals between updates, $2T_c\leq\bar d_j\leq D_{\max}$ provides time for observation and dispatch and bounds the required look-ahead.

The predictor supplies $\hat p_{j,k}(\tau)$ as a function of elapsed execution time $\tau$, with zero power before launch and after completion. A launch at $s_j$ contributes $\hat p_{j,k}(t-s_j)$ at calendar time $t$. Changing $s_j$ shifts the entire trajectory without changing its duration or execution characteristics.

\subsection{Power Aggregation and Phase Adjustment}

Let $p_n^0$ include background demand and the predicted power of running and committed jobs at $t_k+n\Delta_p$. The aggregate prediction over $N$ samples is
\begin{equation}
p_n=p_n^0+\sum_{j\in\mathcal J_k}\hat p_{j,k}(t_k+n\Delta_p-s_j),
\quad n=0,\ldots,N-1.
\label{eq:aggregate}
\end{equation}

The smoothing mechanism can be illustrated by two periodic components with amplitude $A$, frequency $f$, and initial phase difference $\phi$. Delaying the second by $d$ changes their relative phase by $-2\pi fd$. Applying the cosine sum-to-product and half-angle identities yields the aggregate amplitude
\begin{equation}
A_{\mathrm{agg}}^2(d)
=2A^2\left[1+\cos(\phi-2\pi fd)\right].
\label{eq:phase}
\end{equation}

Thus, appropriate launch delays can reduce or cancel coincident power oscillations. The aggregate model in \eqref{eq:aggregate} further accommodates heterogeneous workload profiles, including multiple frequency components and startup variations, without requiring a sinusoidal approximation.

\subsection{Candidate Launches and Evaluation Window}

The feasible launch interval is discretized at resolution $\Delta_d$. Candidate indices $\mathcal M_{j,k}$ and launch times $s_{jm}$ are
\begin{equation}
\mathcal M_{j,k}=
\left\{0,\ldots,
\left\lfloor\frac{\ell_j-t_{k+1}}{\Delta_d}\right\rfloor
\right\},
\quad
s_{jm}=t_{k+1}+m\Delta_d.
\label{eq:candidates}
\end{equation}

For each candidate $m\in\mathcal M_{j,k}$, the predicted power trajectory of job $j$ is calculated as
\begin{equation}
[\mathbf g_{jm}]_n=
\hat p_{j,k}(t_k+n\Delta_p-s_{jm}),
\quad n=0,\ldots,N-1
\label{eq:columns}
\end{equation}
where $\mathbf g_{jm}$ denotes the $N$-sample power trajectory associated with candidate launch time $s_{jm}$.

A binary variable $y_{jm}$ is introduced to indicate whether candidate $m$ is selected for job $j$. Each adjustable job must select exactly one launch time:
\begin{equation}
\sum_{m\in\mathcal M_{j,k}}y_{jm}=1,\quad
y_{jm}\in\{0,1\},\quad j\in\mathcal J_k.
\label{eq:onehot}
\end{equation}

Let $\mathbf p$ and $\mathbf p^0$ denote the aggregate and fixed power vectors over the evaluation window. Stacking the candidate trajectories $\mathbf g_{jm}$ in $A_k$ and the corresponding binary variables in $\mathbf y$ gives the aggregate power and waiting times:
\begin{equation}
\mathbf p=\mathbf p^0+A_k\mathbf y,~
d_j=\sum_{m\in\mathcal M_{j,k}}(s_{jm}-a_j)y_{jm},~
d_j \leq \bar d_j.
\label{eq:affine}
\end{equation}

The first $T_c/\Delta_p$ rows of $A_k$ are zero because adjustable jobs cannot launch before $t_{k+1}$. Aggregate power is affine in the launch decisions.

To evaluate the power fluctuations following even the latest admissible launch, the horizon satisfies
\begin{equation}
(N-1)\Delta_p\geq D_{\max}+T_{\mathrm{tail}}
\label{eq:tail}
\end{equation}
where $T_{\mathrm{tail}}$ covers several cycles of the lowest frequency of interest. This reduces the artificial benefit of postponing fluctuations beyond the horizon.

\section{Optimization Formulation for Start-Time Scheduling}
\label{sec:mpc}
\begin{figure}[t]
\centering
\includegraphics[width=0.48\textwidth]{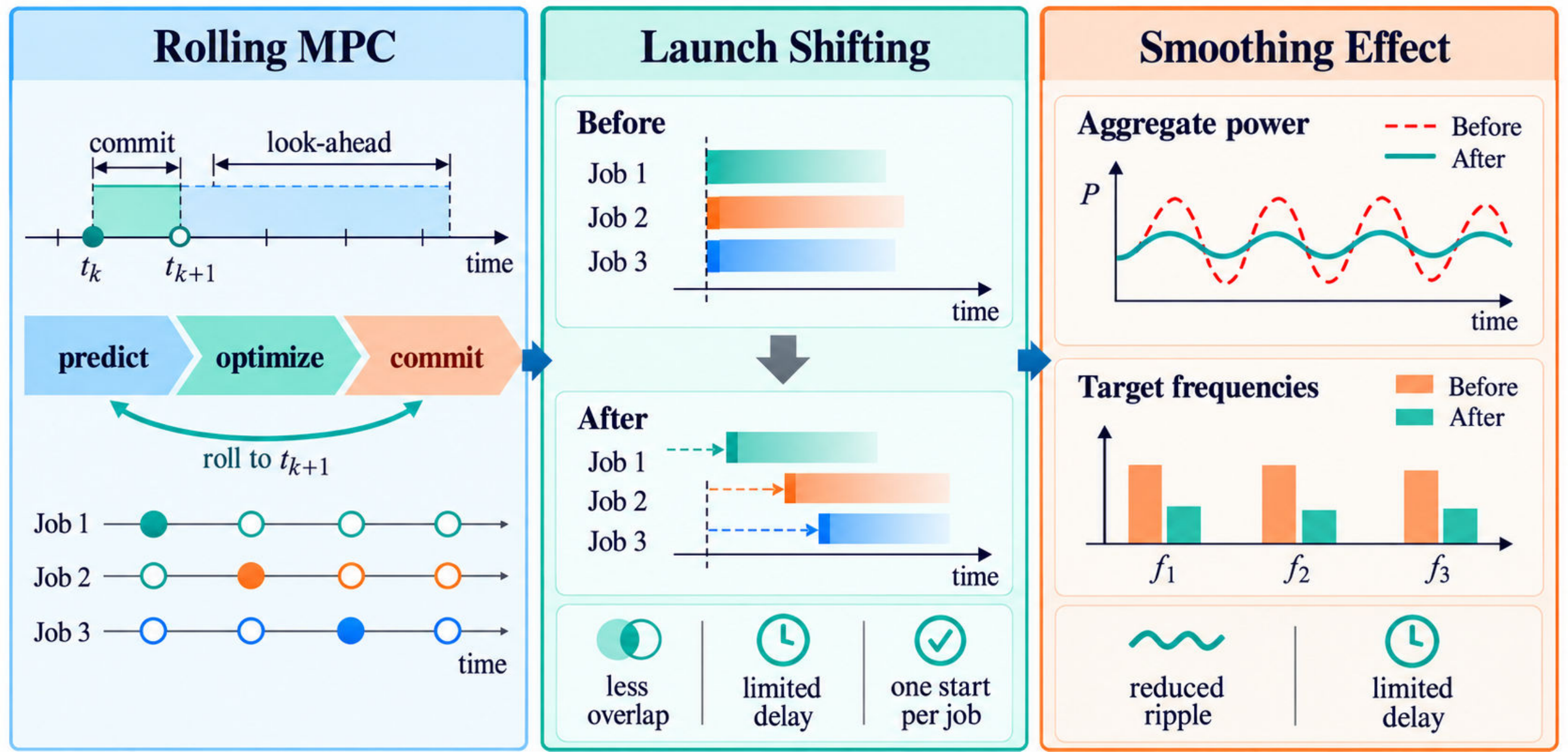}
\caption{Overview of the proposed rolling MPC framework.}
\label{fig:MPC_frame}
\end{figure}

As illustrated in Fig.~\ref{fig:MPC_frame}, the proposed framework repeatedly optimizes job start times. Bounded launch delays coordinate workload phases to suppress aggregate power oscillations. This section formulates the corresponding spectral smoothing objective and optimization model.

\subsection{Spectral Power Smoothing Objective}
Based on the aggregate power model in \eqref{eq:affine}, the proposed scheduling optimization adjusts the launch times of waiting jobs to suppress periodic power fluctuations through workload phase coordination. At each MPC update, a set of dominant frequency bins $\mathcal F_k$ is identified from the predicted workload profiles. For each $q\in\mathcal F_k$, the corresponding frequency is $f_q=q/(N\Delta_p)$, where $N$ is the number of prediction samples.

To quantify the oscillatory component at $f_q$, the aggregate power is projected onto the corresponding cosine and sine basis functions through normalized inner products:
\begin{equation}
\begin{aligned}
u_q&=\frac{1}{N}\sum_{n=0}^{N-1}(p_n-\bar p)
\cos\left(\frac{2\pi qn}{N}\right)\\
v_q&=\frac{1}{N}\sum_{n=0}^{N-1}(p_n-\bar p)
\sin\left(\frac{2\pi qn}{N}\right)
\end{aligned}
\label{eq:projections}
\end{equation}
where $\bar p=N^{-1}\sum_{n=0}^{N-1}p_n$ is the average power, and $u_q$ and $v_q$ are the cosine and sine projection coefficients, respectively.

To establish their relationship with oscillation amplitude, consider a sinusoidal component $x_{q,n}=A_q\cos(\theta_{q,n}+\phi_q)$, where $\theta_{q,n}=2\pi qn/N$, $A_q$ is its amplitude, and $\phi_q$ is its phase. Using the angle-addition identity, the component is expanded as $x_{q,n}=A_q\cos\phi_q\cos\theta_{q,n}-A_q\sin\phi_q\sin\theta_{q,n}$. Since the two basis functions are orthogonal, $\sum_n\sin\theta_{q,n}\cos\theta_{q,n}=0$, and their squared norms satisfy $\sum_n\cos^2\theta_{q,n}=\sum_n\sin^2\theta_{q,n}=N/2$. Substituting into \eqref{eq:projections} yields
\begin{equation}
\begin{aligned}
&u_q=\frac{A_q\cos\phi_q}{N}\sum_n\cos^2\theta_{q,n}=\frac{A_q}{2}\cos\phi_q,\\
&v_q=-\frac{A_q\sin\phi_q}{N}\sum_n\sin^2\theta_{q,n}=-\frac{A_q}{2}\sin\phi_q,\\
&u_q^2+v_q^2=\frac{A_q^2}{4}.
\end{aligned}
\label{eq:projection_magnitude}
\end{equation}

Thus, $u_q^2+v_q^2$ characterizes the aggregate oscillatory power at frequency $f_q$ independently of its absolute phase. By adjusting job launch times, the relative phases of individual workload components can be coordinated to promote their cancellation. Accordingly, the spectral power-smoothing objective is formulated as
\begin{equation}
J_f=\sum_{q\in\mathcal F_k}\left(u_q^2+v_q^2\right).
\label{eq:spectral}
\end{equation}

\subsection{Optimization Model Formulation and Solution}

Based on the spectral smoothing objective in \eqref{eq:spectral}, the launch times of waiting jobs are optimized to minimize aggregate power fluctuations while avoiding unnecessary scheduling delays. For each candidate launch time $s_{jm}$, the normalized squared waiting cost is defined as $c_{jm}^{d}=[(s_{jm}-a_j)/\bar d_j]^2$. Using the binary launch decisions in \eqref{eq:onehot} and the aggregate power model in \eqref{eq:affine}, the optimization problem at each MPC update is formulated as
\begin{equation}
\begin{aligned}
\min_{\mathbf y,\mathbf p}\quad J={}&
J_f +\frac{\lambda_d}{|\mathcal J_k|}
\sum_{j\in\mathcal J_k}
\sum_{m\in\mathcal M_{j,k}}
c_{jm}^{d}y_{jm}\\
\mathrm{s.t.}\quad&
\eqref{eq:onehot},\ \eqref{eq:affine}
\end{aligned}
\label{eq:optimization}
\end{equation}
where $\lambda_d>0$ controls the relative importance of waiting time. The waiting cost increases quadratically with the fraction of the allowable delay, discouraging excessive postponement. The normalization by $|\mathcal J_k|$ accounts for variations in the number of waiting jobs.

Since the spectral coefficients $u_q$ and $v_q$ in \eqref{eq:projections} are linear in the aggregate power $\mathbf p$, and $\mathbf p$ is affine in the binary launch decisions $\mathbf y$, the spectral objective is a convex quadratic function of $\mathbf y$. Meanwhile, the waiting-cost coefficients $c_{jm}^{d}$ are precomputed, making the additional penalty linear in $\mathbf y$. The resulting optimization model can therefore be solved using standard mixed-integer optimization solvers such as Gurobi. Precomputing the candidate power trajectories further avoids repeated workload-profile calculations during optimization.

\subsection{Online Scheduling Implementation}

The proposed start-time scheduling method is implemented in a receding-horizon manner, as summarized in Algorithm~\ref{alg:rolling_scheduling}. At each MPC update, the optimization model is solved using the latest workload power predictions. To ensure timely execution, a feasible schedule is maintained by retaining previous uncommitted launch decisions and assigning newly arrived jobs to their earliest admissible launch times. The optimization is performed within a prescribed computation budget, and the best available feasible solution is adopted. Under the scheduling conditions established in Section \ref{sec: MPC_framework}, this procedure preserves launch-time feasibility even when the optimization terminates before reaching optimality.

\begin{algorithm}[htbp]
\caption{Online Start-Time Scheduling}
\label{alg:rolling_scheduling}
\KwIn{Job information, power predictions, and previous provisional schedule}
\KwOut{Committed launch decisions and updated provisional schedule}
\SetAlgoLined

\For{each MPC update $t_k$}{
Update job states and predicted power profiles\;

\If{$\mathcal J_k\neq\emptyset$}{
Construct candidate launch trajectories and the aggregate power model using \eqref{eq:candidates}--\eqref{eq:affine}\;

Identify the dominant frequency set $\mathcal F_k$ and construct the spectral objective \eqref{eq:spectral}\;

Initialize with retained launch decisions and earliest feasible starts for new jobs\;

Solve \eqref{eq:optimization} within the prescribed computation budget\;

Keep the best feasible plan available by the dispatch deadline\;

}
Commit launch decisions within $[t_{k+1},t_{k+2})$\;

Retain subsequent launch decisions\;
}
\end{algorithm}

\section{Case Study}
\subsection{Case Setup}
The case study considers a DC hosting training and fine-tuning jobs with heterogeneous power demands and execution durations. Workload profiles follow the alternating computation and communication phases in \cite{11487948}, with heterogeneous power levels and cycle-to-cycle variations. Training cycles are near 1~Hz, whereas fine-tuning frequencies span 0.3--0.7~Hz. The controller adopts a power-sampling interval of $\Delta_p=0.05$~s, a launch-time resolution of $\Delta_d=0.10$~s, an MPC update period of $T_c=0.50$~s, and a 30-s prediction horizon. All scheduling decisions follow a common absolute time grid, with a maximum waiting time of $\bar d_j=D_{\max}=4$~s. At each MPC update, dominant frequencies are identified from both the predicted aggregate load and individual waiting-job profiles. Within 0.2--2~Hz, up to six dominant peaks and their adjacent frequency bins are included in $\mathcal F_k$. The model is implemented in Pyomo and solved with SCIP, using feasible local-search initialization and a 0.05 s solver budget. 

To evaluate the proposed method, three scheduling strategies are compared: \emph{M1} launches each job at the earliest admissible time; \emph{M2 } uniformly selects a feasible launch time and retains it until launch; and \emph{M3 (proposed scheduling)} optimizes launch times using the phase-aware MPC framework. All methods use identical job profiles and 280--580~s observation interval.

\subsection{Power Smoothing Through Phase Coordination}
\begin{figure}[t]
\centering
\includegraphics[width=0.45\textwidth]{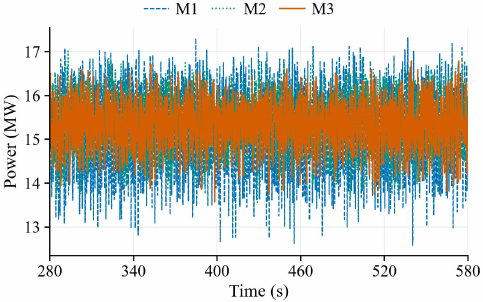}
\caption{Aggregate DC power profile.}
\label{fig:case_power}
\end{figure}

Fig.~\ref{fig:case_power} shows that M2 and M3 narrow the fluctuation envelope compared with M1, while maintaining an average demand of approximately 15.35~MW. The reduced excursions result from improved temporal alignment between the computation and communication phases of different jobs, with individual power profiles preserved. Table~\ref{tab:case_metrics} quantifies this improvement in both the frequency and time domains. M3 achieves the lowest band mean square, power standard deviation, and peak-to-peak variation, reducing these measures by 22.35\%, 9.97\%, and 6.69\% relative to M2, respectively. These results show that phase-aware launch selection further suppresses both periodic oscillations and aggregate power excursions beyond the smoothing provided by random staggering.

\begin{table}[t]
\caption{Comparison of Power Fluctuations}
\label{tab:case_metrics}
\centering
\begin{tabular}{cccc}
\hline
Metric & M1 & M2 & M3\\
\hline
0.2--2~Hz mean square (MW$^2$) & 0.79144 & 0.26099 & 0.20266\\
Power standard deviation (MW) & 0.92619 & 0.54938 & 0.49458\\
Peak-to-peak power (MW) & 4.73431 & 3.45345 & 3.22258\\
\hline
\end{tabular}
\end{table}

\begin{figure}[t]
\centering
\includegraphics[width=0.45\textwidth]{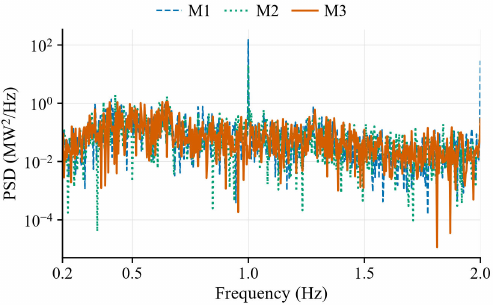}
\caption{Power spectra over the complete observation interval.}
\label{fig:case_spectrum}
\end{figure}

The spectra in Fig.~\ref{fig:case_spectrum} explain this difference. M1 leaves the nearly 1-Hz training cycles closely aligned. M2 disperses their phases but does not explicitly minimize the remaining aggregate oscillation. M3 instead selects delays jointly, accounting for both incoming jobs and the fixed profiles of running jobs. A relative delay of 0.5~s reverses the phase of a 1-Hz component; suitable offsets therefore allow individual fluctuations to partially cancel. The residual 1-Hz peak under M2 is further attenuated by M3. Heterogeneous cycles still produce broadband fluctuations that cannot all be canceled by fixed launch offsets.

The band mean squares in Table~\ref{tab:case_metrics} are obtained by integrating the spectra over 0.2--2~Hz. Relative to M1, M2 reduces this measure by 67.02\%, whereas M3 achieves a 74.39\% reduction. Thus, random staggering provides substantial smoothing, while phase coordination further suppresses the residual oscillation. Random staggering does not necessarily balance the oscillatory contributions of jobs with different power levels and cycle frequencies. M3 coordinates their launch offsets to improve cancellation among dominant components while preserving individual power profiles.

\begin{figure}[t]
\centering
\includegraphics[width=0.45\textwidth]{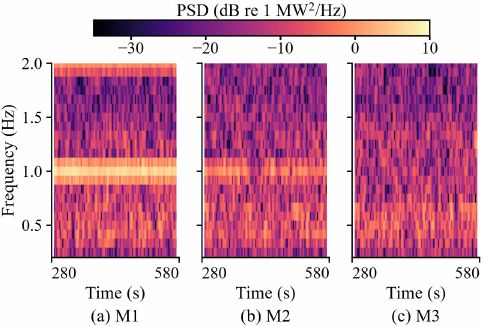}
\caption{Time-frequency power spectra.}
\label{fig:case_time_frequency}
\end{figure}

Fig.~\ref{fig:case_time_frequency} compares all three methods. The strong 1-Hz band under M1 is reduced by M2 and further weakened by M3 over most of the interval. All jobs have already started; the sustained difference therefore results from launch offsets, without interrupting or retiming running jobs.

\subsection{Waiting Cost and Computational Performance}
\begin{figure}[t]
\centering
\includegraphics[width=0.45\textwidth]{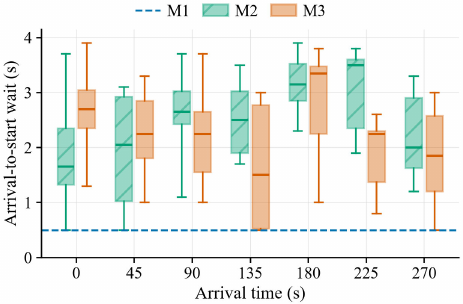}
\caption{Arrival-to-start waits by batch.}
\label{fig:case_delays}
\end{figure}

Fig.~\ref{fig:case_delays} compares arrival-to-start waits by batch. Mean waits are 2.476 and 2.188~s for M2 and M3. Both delayed-launch methods have a maximum wait of 3.9~s and satisfy the 4-s limit. The benefit comes from selecting suitable relative offsets rather than simply increasing delays. All jobs retain their original execution durations and complete-job energy consumption. For each MPC update, total computation, including initialization and optimization, takes a median of 0.216~s and a maximum of 0.390~s, within the 0.50-s control period.

\section{Conclusion}
This paper proposed an MPC-based start-time scheduling method for smoothing AI data-center power. Bounded launch delays change the relative phases of workload oscillations while preserving each job's uninterrupted execution, providing a software-based smoothing mechanism without additional storage equipment. In the continuous-operation case, phase coordination suppresses the dominant spectral peaks and reduces the 0.2--2-Hz power mean square by 74.39\% and the power standard deviation by 46.6\%. These improvements require only 1.688~s of additional mean waiting relative to earliest start, with all launch deadlines met and every update completed within 0.50~s. Future work will examine the effect of forecast errors and computing-resource constraints.

\bibliographystyle{IEEEtran}
\bibliography{ref}

\end{document}